\documentclass[conference, letterpaper]{IEEEtran}

\usepackage[OT1]{fontenc}

\usepackage[noadjust]{cite}

\usepackage{url}

\usepackage{amsmath}
\usepackage{graphicx}
\usepackage{booktabs}
\usepackage{multirow}
\usepackage{enumitem}
\usepackage{tabularx}
\usepackage{balance}
\usepackage{placeins}
\usepackage{afterpage}
\usepackage{threeparttable}

\DeclareMathOperator{\TTP}{TTP}
\DeclareMathOperator{\TTE}{TTE}

\usepackage{fancyhdr}
\begin{document}

\title{A Bug Bounty Perspective on the \\ Disclosure of Web Vulnerabilities}

\author{
\IEEEauthorblockN{Jukka Ruohonen}
\IEEEauthorblockA{University of Turku \\ 
Email: juanruo@utu.fi}
\and
\IEEEauthorblockN{Luca Allodi}
\IEEEauthorblockA{Eindhoven University of Technology \\
Email: l.allodi@tue.nl}
}

\maketitle

\begin{abstract}
Bug bounties have become increasingly popular in recent years. This paper discusses bug bounties by framing these theoretically against so-called platform economy. Empirically the interest is on the disclosure of web vulnerabilities through the Open Bug Bounty (OBB) platform between 2015 and late 2017. According to the empirical results based on a dataset covering nearly 160 thousand web vulnerabilities, (i) OBB has been successful as a community-based platform for the dissemination of web vulnerabilities. The platform has also attracted many productive hackers, (ii) but there exists a large productivity gap, which likely relates to (iii) a knowledge gap and the use of automated tools for web vulnerability discovery. While the platform (iv) has been exceptionally fast to evaluate new vulnerability submissions, (v)~the patching times of the web vulnerabilities disseminated have been long. With these empirical results and the accompanying theoretical discussion, the paper contributes to the small but rapidly growing amount of research on bug bounties. In addition, the paper makes a practical contribution by discussing the business models behind bug bounties from the viewpoints of platforms, ecosystems, and vulnerability markets.
\end{abstract}

\begin{IEEEkeywords}
vulnerability disclosure, vulnerability reward program, bug hunting, bug challenge, open bug bounty, security patching, web vulnerability, cross-site scripting, XSS, CSRF
\end{IEEEkeywords}

\section{Introduction}

Bug bounties have become increasingly popular in recent years. As a testimony of the popularity, even the United States Department of Defense (DoD) recently piloted a bug bounty program, which further led to a partnership with a crowd-sourcing bug bounty platform~\cite{Chatfield17, Kinis17}. Despite of the popularity, bug bounties are surrounded by many unanswered and controversial questions. These questions range from monetary incentives and ethical practices to the fundamental question of whether bug bounties actually help at improving security.

Motivated by the many unanswered questions, this paper examines the vulnerability disclosure dynamics on the one-sided OBB platform that was launched in 2015 based on the older volunteer-driven XSSPosed platform \cite{SCMedia15}.\footnote{~The paper covers a period from the platform's initial launch to late 2017. As is typical to current bug bounty platforms, the underlying business models are constantly changing~\cite{Hata17, Ring14}. For instance, OBB recently started to gear itself toward managing vendors' bug bounty programs, while also permitting the submission of new types of web vulnerabilities. Despite of these changes,  most of what is being discussed apply also to the situation in mid-2018.} The term vulnerability \textit{disclosure} frames the scope of the paper: the primary focus is on the dissemination of vulnerabilities through the OBB platform to the vendors affected by the vulnerabilities. The term \textit{one-sided} is used to emphasize that OBB is mostly a community-based platform that neither pays for the vulnerabilities disseminated nor explicitly engages with vendors through a subscription model. As will be shown, both terms are important for a theoretical framing of bug bounties.

The paper's main empirical findings, theoretical points, and contributions can be summarized and generalized as follows:
\begin{itemize}
\vspace{1pt}
\itemsep0.5em 
\item{Bug bounty platforms in general align well with the \textit{theories about network effects and platform economy}.}
\item{However, when \textit{excluding the crowd-sourcing} elements, \textit{innovation seems limited} from a business perspective; most current bug bounties \textit{mimic the business models that have been used already in the older vulnerability markets}.}
\item{\textit{One-sided bug bounty platforms for web vulnerabilities represent an interesting case of comparison} to two-sided bug bounty platforms such as HackerOne and the older platforms such as the notorious Zero Day Initiative (ZDI).}
\item{In terms of the mere volume of software vulnerabilities disseminated, bug bounties \textit{can be successful without monetary compensations}, although the lack of compensations tends to intensify the focus on \textit{quantity over quality}.}
\item{Only a relatively \textit{few participants disclose most} of the web vulnerabilities disseminated through bug bounties.}
\item{\textit{Automated tools for web vulnerability discovery are used} also in the bug bounty context, and this automation presumably \textit{influences the websites targeted and affected}.}
\item{In addition to the productivity gap between participants, the use of automated tools can create \textit{knowledge gaps} even with respect to a single type of web vulnerabilities.}
\item{The \textit{evaluation of new submissions can be rapid} in the context of simple web vulnerabilities, although it remains an \textit{open question of how well the vulnerabilities disseminated are coordinated and communicated to vendors}.}
\item{\textit{Patching of vulnerabilities by the vendors affected takes a relatively long time} also in terms of low-impact web vulnerabilities often disseminated through bug bounties.}
\item{Patching times vary across both participants and vendors, but \textit{learning from the past disseminated vulnerabilities seems limited}; the reputation of a bug bounty is likely a more important factor affecting the patching times.}
\vspace{2pt}
\end{itemize}

Three additional remarks are required about the terms used.  First and foremost: there exists no established terminology to describe the current crowd-sourcing patterns for vulnerability discovery and disclosure. Bug bounties, vulnerability reward programs, security challenges, vulnerability hunting campaigns, and related terms are used more or less interchangeably to describe the same phenomenon \cite{KuehnMueller14b}. Throughout this paper, the term \textit{bug bounty} is used in the same loose sense. Second, the term (white hat) \textit{hacker} is used for referring to individual participants in bug bounties. Because particularly criminology research has started to equate hackers with computer crime~\cite{Holt10}, the term security researcher often used in the industry would be a slightly better choice. Nevertheless, both terminology choices are justifiable due to the concrete connections to the current security industry---the term bounty appears in the OBB abbreviation and the term hacker in the name of the likely most famous current bug bounty platform, HackerOne. Last but not least, the term \textit{vendor} is often used to refer to ``any producer of software, regardless of whether or not that software is sold commercially''~\cite{Ozment07}. In this paper, however, vendors are equated to domain names that have hosted the websites affected by the vulnerabilities observed. 

The last point requires a further comment. The owner of a domain may not be the same party who is responsible for the website(s) hosted from the domain. In fact, the remediation of some web vulnerabilities may involve domain name registrants, webmasters, hosting providers, and even Internet service providers~\cite{Tajalizadehkhoob17}. In theory, the same may apply to the disclosure of web vulnerabilities. As the bug bounty platform examined does not provide information about the actual vendor-side individuals who were contacted about the vulnerabilities, domain names provide a sensible simplification, however. As will be elaborated, questions related to contact persons differentiate also many bug bounty platforms.

Bug bounties are a challenging research topic. There is at the same time a limited but growing amount of existing research to build upon~\cite{Fryer17}, and a large amount of loosely related research on vulnerabilities. In terms of scholarly disciplines, relevant contributions have been made in computer science and software engineering, information systems research, and what has been branded as (cyber) security economics~\cite{AndersonMoore09}. In order to maintain this interdisciplinary focus, the remainder of this paper proceeds by first discussing the background in Section~\ref{section: background} from a socio-technical perspective. This theoretical discussion is used to also motivate the empirical case study presented in Section~\ref{section: results}. The empirical findings, theoretical points, and few practical insights are discussed in the final Section~\ref{section: discussion}.

\section{Background}\label{section: background}

The following discussion will outline the background by considering some similarities and differences between current bug bounties and the older vulnerability markets. Even though formal economic models have been proposed for approaching bug bounties~\cite{ChoiFershtman10}, the theoretical tone adopted for the discussion is more informal, drawing from the platform literature.

\subsection{Bug Bounties}

\begin{figure}[th!b]
\centering
\includegraphics[width=8cm, height=9.3cm]{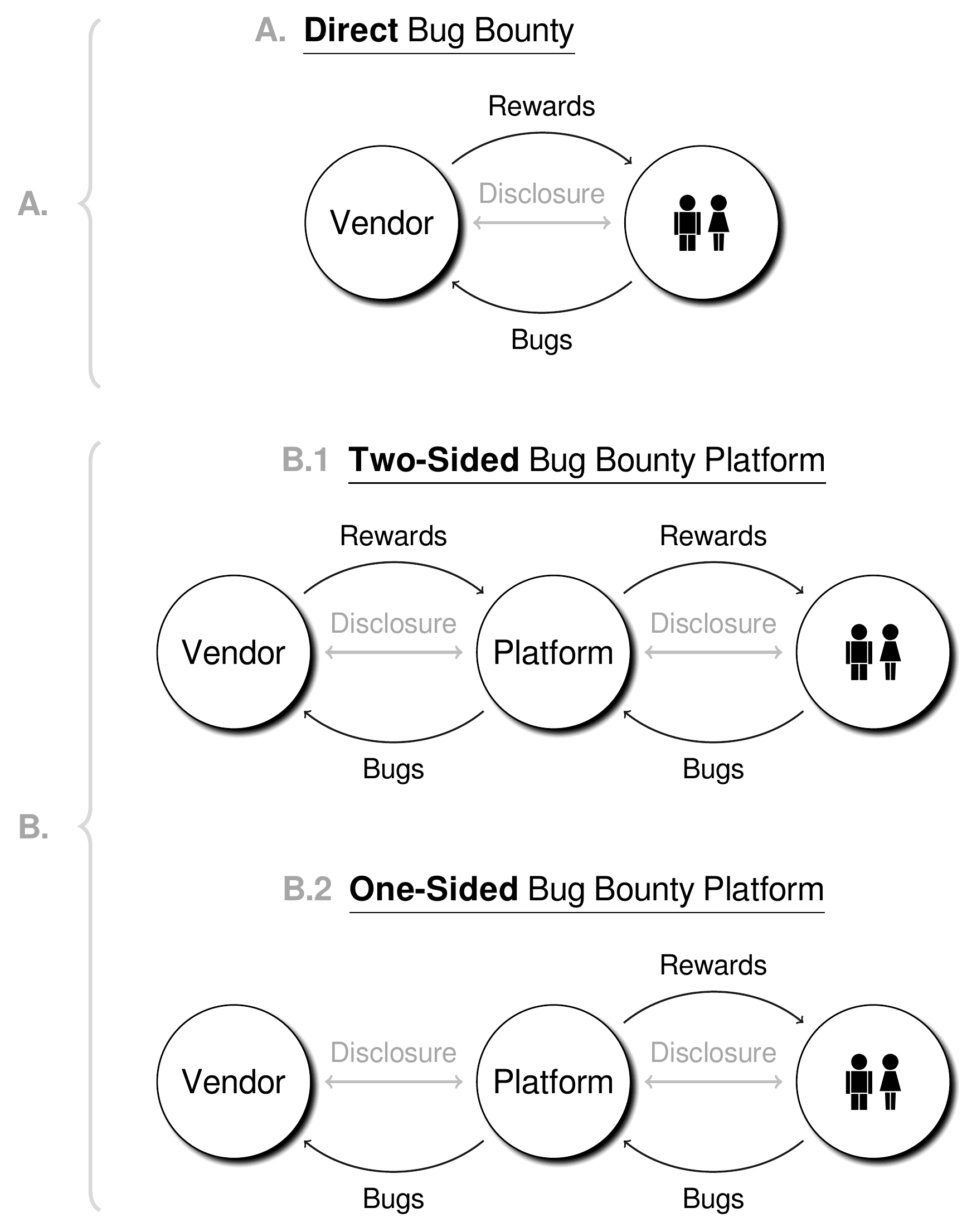}
\caption{Bug Bounty Variants}
\label{fig: bug bounty}
\end{figure}

The history of bug bounties traces to the early 2000s emergence of commercial vulnerability disclosure programs and different security alerting services. The perhaps most memorable example is the TippingPoint's Zero Day Initiative that was launched already in 2005. This still active program relies on a business model that compensates hackers for their vulnerability discoveries on one hand, and helps the opt-in customers to patch their products on the other \cite{AndersonMoore09}. When reflected against the history of vulnerability disclosure practices, the new element brought forward by ZDI and related programs was the monetary compensations paid for the vulnerabilities disseminated through the programs. These monetary compensations were also a key element in the emergence of crowd-sourced bug bounty programs in the early 2010s. There are currently two main variants of bug bounty programs~\cite{Fryer17}. These are illustrated in Fig.~\ref{fig: bug bounty}. As a preparation for the forthcoming theoretical discussion, the second variant (B) is further broken down into two subvariants, B.1 and B.2.

Direct bug bounty programs are nowadays orchestrated by many vendors themselves. In fact, the concept of a (direct) bug bounty is older than ZDI and related programs; Netscape introduced the first known bug bounty already in 1995. Although also the Mozilla Foundation later adopted the same approach, these early initiatives did not gain wide\-spread traction in the software industry. It was much later in the early 2010s when direct bug bounties became commonplace through the initiation of programs by many technology giants, including Google, Microsoft, Facebook, and Yahoo, among others. These bug finding contests directly orchestrated by technology companies themselves were also quickly adopted for initiating different platforms (see~\cite{Facin16} for the concept of platform and related theory). HackerOne is currently the perhaps most famous case of these bug bounty platforms.  

\subsection{Bug Bounty Platforms}

Many of the two-sided platforms (B.1) rely on modified versions of the business models that were used already in ZDI and related programs. For instance, vendors subscribe to HackerOne in order to improve the security of their software products via security assessments carried out by hackers who are compensated for their vulnerability discoveries. As with ZDI and some later online marketplace endeavors~\cite{Ruohonen16RCIS}, the platform exploits a two-sided market; the platform enables two distinct groups to interact and transact (security) information according to their distinct needs~\cite{Voigt16}. Unlike the vulnerability and security data feeds provided in ZDI and related programs, however, HackerOne is a proactive rather than a reactive service; vendors are explicitly exposing their products for security assessments. For vendors, the service provided could be even labeled as crowd-sourced penetration testing. 

It is important to further remark that there is no need for vendors to participate in order to avoid missing vulnerability information. Consequently, many but not all \cite{ZhaoGrossklags15} of the business models lack the element of blackmailing that is implicitly present in the ZDI-style programs~\cite{Bohme06}. Because vendors are the paying customers, nevertheless, the revenue streams are still comparable to those used by the older vulnerability brokers.

To some extent, the demarcation between the two main theoretical types (A and B) has become slightly blurry as some vendors have started to modify their direct bug bounties toward the direction of platforms. For instance, Google extended its bug bounty program in 2013 to cover also a few security-critical open source software projects \cite{Zalewski13}. Analogously, a few years later in 2016 the Mozilla Foundation launched a fund to improve security in the open source domain~\cite{Riley16}. This initiative can be also considered a platform because the fund is used to pay for security audits and vulnerability discoveries, which are both coordinated through Mozilla. Moreover, bounty systems have recently been expanded toward areas beyond security. Possibly inspired by bug bounties and vulnerability hunting, these two-sided crowd-sourcing platforms offer monetary bounties for implementing new features and fixing conventional (non-security) bugs~\cite{GaoWang18, Kanda17}. A further trend relates to the arrival of different one-sided bug bounty platforms, including the OBB platform studied later on in Section~\ref{section: results}.

Unlike the two-sided subvariants (B.1), these one-sided platforms do not seek to monetize vendor involvement. Although vendors are encouraged to provide voluntary compensations~\cite{OBB17}, the one-sided platforms (B.2) do not explicitly pay for the vulnerabilities reported. In this sense, these platforms are not pure vulnerability marketplaces on which each vulnerability is a unit of trade~\cite{McKinney07}. Instead, the one-sided subvariants provide community-based platforms for hackers to report and disclose vulnerabilities they have discovered. These characteristics imply that the ethos behind the one-sided platforms is closer to the classical topics in vulnerability disclosure. The absence of compelled monetary compensations also implies that the rewards from participation are intrinsic rather than extrinsic. Before proceeding to discuss these rewards in more detail, it should be briefly noted that the two subvariants share both similarities and dissimilarities in terms of the basic theoretical premises for platform success.

\subsection{Network Effects}

A fundamental challenge for any platform orchestrator has always related to the creation and maintenance of a critical mass~\text{\cite{OndrusLyytinen15, Ruutu17}}. This question is also well-understood among the current bug bounty platform orchestrators. For instance, many bug bounty websites market themselves by visible announcements about the amount of vendors and hackers who have participated, the amount of vulnerabilities reported through the platform in question, the amount of compensations paid, and so forth. These marketing techniques relate to the concept of network effects~\cite{KatzShapiro85}, which can be further chopped analytically into cross-side (or indirect) network effects and direct (or one-side) network effects \cite{Haile16, OndrusLyytinen15, Ruutu17}. The former effects mean that the value of participating on one side of a platform depends on the amount of participants on the other side. Thus, the more there are hackers participating on a two-sided bug bounty platform, the more there are incentives for vendors to also participate, and the other way around. 

Similarly to dating platforms \cite{Voigt16}, say, these cross-side network effects are also numerically asymmetric in two-sided bug bounty platforms. For a platform orchestrator, one high-profile vendor (such as Adobe or Intel---or DoD) is worth a hundred hackers, in a manner of speaking. As the one-sided platforms do not require explicit vendor participation on the platforms, these cross-side network effects pose a bigger challenge only for the two-sided bug bounty platforms. 

The direct network effects refer to theoretical premises in which the amount of participants on one side influences the value for participation on this same side. Social media platforms would be the prime example of such direct network effects; there is only a small incentive to participate on a social media platform unless there are already plenty of other participants. For the two subvariants of bug bounty platforms, there exist analogous challenges related to direct network effects---a critical mass of hackers is required. Likewise, a large  amount of well-known vendors participating on a two-sided bug bounty platform likely increases the likelihood that also other vendors will join. Reputation and trust offer one way to meet these requirements in the context of online markets for vulnerabilities and related security items~\cite{AllodiMassacci16, Miller07, Ruohonen16RCIS}. Different rewards for the vulnerabilities reported provide another way to attain and maintain a critical mass of participants.

\subsection{Rewards}\label{subsec: rewards}

The concept of reward is important for better understanding bug bounty platforms in particular. Five points are worth making about the concept of reward in the bug bounty context. 

First, there are monetary rewards and non-monetary rewards~\cite{KuehnMueller14b}. Monetary rewards are the ones grabbing the attention in popular discourse---the outlying payments that can amount even up to a hundred thousand dollars make catchy headlines in media. Although empirical research about the money involved is regrettably limited, there are good reasons to suspect that most monetary compensations are quite moderate, however (see  \cite{Allodi17c, Finifter13, Ransbotham12, Ruohonen16RCIS, ZhaoGrossklags15}, though note also~\cite{HackerOne18a}). Despite of the popularity of bug bounties, it seems fair to also question whether a stable supply-demand equilibrium has been reached in terms of the monetary compensations. The history provides a rationale for this caution: the lack of clear reference prices has been a continuing problem in the vulnerability markets \cite{Miller07, Ruohonen16RCIS}. Furthermore, recent industry surveys also indicate that monetary compensations are not commonly expected from reporting vulnerabilities~\cite{Uchil16}. Therefore, it is important to emphasize the role of non-monetary rewards. 

The non-monetary rewards include everything from free-lunch parties to \text{t-shirts}, but the most important factor relates to the ``acknowledgments in the security hall of fame of the respective bounty program''~\cite{KuehnMueller14b}. Given that career development is one intrinsic motivation for participating in open source software development~\cite{Crowston12}, having a name in a security hall of fame may be a good abstract business card when seeking employment in security companies~\cite{Ring14}. For this reason, it should be noted that the importance of acknowledgements is not limited only to bug bounty platforms; also many companies that do not engage with bug bounties maintain their own halls of fame. For instance, the Finnish company Nokia maintains its own hall of fame for those who have disclosed vulnerabilities in the company's products~\cite{Nokia17}, although the company currently neither maintains a direct bug bounty nor participates in two-sided platforms as a customer.

Second, there are non-monetary rewards that originate from the collaboration between hackers participating on a platform. When diverse members of a cohesive social group interact, the shared interests typically facilitate knowledge sharing among the members of the group~\cite{MendezDuron09, Olaisen17}. As bug bounty platforms help at networking with colleagues \cite{Ring14}, these may also increase the competency of a member either through collaboration with other members or by learning from others~\cite{ZhaoGrossklags14}. By implication, such 
rewards are also related to direct network effects; the larger the amount of participants, the more there are opportunities for collaboration and learning. As is typical to open source software development and online communities in general~\cite{LiTanTeo12, ZhangLiuDengChen17}, also different intrinsic incentives are present~\cite{Ring14}. These include abstract rewards related to social approval, enjoyment and leisure time, intellectual stimulation, and other sociological and psychological aspects.

Third, there are rewards for hackers, and costs and rewards for vendors. Orchestrating and maintaining a direct bug bounty program is not free. Depending on a vendor's software portfolio and its size, already the maintenance costs can be noteworthy. These and other related costs are an important element in the business models behind many two-sided bug bounty platforms. For vendors seeking to outsource a portion of security assessments to a crowd, it is presumably cheaper and easier to organize the outsourcing via a third-party platform. Liability, communication, and related aspects likely further increase the lucrativeness of two-sided platforms. 

Unfortunately, no empirical research has been done to examine the pricing of bug bounty platforms for vendors. It can be noted that measuring the costs may not be easy because there may be indirect costs in addition to the direct participation expenses. For instance, a vendor who participates in a bug bounty platform may be exposed to a deceptive incentive to underinvest in other secure software engineering practices~\cite{Lam16}, possibly due to the perceived cost-effectiveness of bug bounties~\cite{Finifter13, Such16}. That is, paying for vulnerabilities may diminish resources from the prevention of vulnerabilities~\cite{Egelman13}. There is also lack of research on the rewards that two-sided bug bounty platforms offer for vendors. To pinpoint directions for further research in this regard, it can be noted that besides security assurance itself, marketing and public relations  constitute a reward. When a vendor participates in a popular and widely known two-sided platform such as HackerOne, it also delivers a public statement that security is taken seriously---regardless whether this is actually true. 

Fourth, most current bug bounty platforms harness the rewards for making their platforms profitable and lucrative for both hackers and vendors. Taking cuts from transactions was already a part of the older vulnerability brokerage models (currently, HackerOne takes 20\% of the monetary rewards offered by vendors~\cite{Hackerone17b}). The harnessing extends also toward the extrinsic and intrinsic rewards for the hackers. For achieving and maintaining the critical mass, many bug bounty platforms rely on so-called gamification techniques (for the concept of gamification see, e.g.,~\cite{Garcia17}). These techniques include metric-based rankings and constantly updated dashboards, badges for most productive hackers, and other commonly used social reputation elements. Moreover, having a name in a platform's security hall of fame is not enough; it is encouraged to also compete in a constantly updated dashboard~\cite{HuangLiu16}. As soon discussed, it should be emphasized that both the extrinsic and intrinsic rewards vary in terms of different vulnerability types.

Last, there exists an implicit societal reward. In theory, bug bounties may reduce exploitable vulnerabilities stockpiled by criminals and state-level actors, providing also pull-off incentives that may decrease the probability of participating on illegal underground platforms \cite{Allodi17c, ZhaoLaszka17}. The direct network effects involved imply that an already reached critical mass may be also educated and guided toward established security practices, ethical codes of conduct, and more sophisticated vulnerabilities. These points resonate with the classical but still ongoing debates about software vulnerability markets in general. Thus, depending on a viewpoint, the societal reward may  also be a liability: instead of working toward the ultimate goal of improving software quality, bug bounties may increase the stockpiling tendency and the exploitation of vulnerabilities~\cite{Chatfield17, KuehnMueller14b, Egelman13}. Given these fundamental problems, it is important to emphasize that most current crowd-sourcing bug bounty platforms target low-impact web vulnerabilities.

\subsection{Disclosure on Bug Bounty Platforms}

The historical genesis behind the ZDI-like programs was largely related to the altruistic motives in the 1990s full disclosure movement~\cite{McKinney07}. These motives can be also seen underneath the current bug bounty platforms. To recall, full disclosure refers to a vulnerability disclosure practice via which full technical details are released to the public, possibly regardless whether a vendor was even contacted about the vulnerabilities prior to the release. There were---and still are---good reasons for hackers to prefer this type of vulnerability disclosure. Among these is the reluctance of many vendors to acknowledge and fix the vulnerabilities reported. An intermediate actor is one way to address this problem and related issues affecting the disclosure of software vulnerabilities.

Direct bug bounty programs share a key similarity with the so-called direct disclosure practice through which hackers and vendors communicate privately about the publication and patching of vulnerabilities. This direct (or two-party) disclosure practice~\cite{CERT17} has historically been the most common way to disclose vulnerabilities. Many of the alternative practices were also formulated to overcome limitations affecting the two-party direct disclosure type. These alternatives include the full disclosure ideology,  disclosure through computer emergency response teams (CERTs), and the ZDI-like brokerage solutions \cite{Kinis17, Ransbotham12}. The similarity between direct disclosure and direct bug bounties relates to the absence of a middleman, whether a commercial vulnerability broker or a CERT. However, there exists also a fundamental difference: with direct bug bounties, vendors are well-prepared to handle vulnerabilities disclosed to them. Whereas a classical direct disclosure of a vulnerability may come out from the blue sky to a vendor, a vulnerability disclosed through a direct bounty program is---or at least should be---received by a dedicated contact team.

The current bug bounty platforms lean toward either the direct disclosure practice or the hybrid brokerage models. A key differentiating factor relates to a platform's role in coordinating the disclosure. If a platform takes a weak brokerage position, the platform does not explicitly coordinate the disclosure process between vendors and hackers; the process cannot be outsourced to the bug bounty platform. For instance, the primary way to handle disclosure on the HackerOne's platform is to directly disclosure information to vendors based on the contact details provided by the platform. While there is an additional service for helping hackers with the initial handshaking \cite{HackerOne17a}, the service does not mean that HackerOne would be the primary coordinator. Analogous point applies to the OBB platform. In contrast, some bug bounties take a much stronger brokerage position, handling all communication with vendors on behalf of the hackers \cite{Ring14}.  It should be noted that some implicit mediation is still present even when explicit brokerage is not implemented. Any bug bounty platform is still implicitly present in the disclosure processes; already the name of a platform may carry some authority for influencing vendors' behavior. When a vendor knows that a bug bounty platform is involved, communication may be easier compared to classical pure direct disclosure. This theoretical reasoning is summarized with the cross-tabulation shown in Table~\ref{tab: brokerage}.

\begin{table}[th!b]
\begin{center}
\caption{A Topology of Disclosure Brokerage}
\label{tab: brokerage}
\begin{tabular}{llccc}
\toprule
& & \multicolumn{2}{c}{Brokerage} \\
\cmidrule{3-4}
& & Weak & Strong \\
\cmidrule{3-4}
\multirow{2}{*}{Platform} & Two-sided & HackerOne & ZDI \\
\cmidrule{2-4}
 & One-sided & OBB & Vulnerability Lab? \\
\bottomrule
\end{tabular}
\end{center}
\end{table}

While different forms of brokerage may be implemented in both one-sided and two-sided platforms, there is a key difference between these two theoretical types with respect to vulnerability disclosure---vendors are explicitly participating on two-sided platforms as paying customers. Due to the lack of explicit vendor-side engagement, vulnerability disclosure is presumably more difficult to carry out through one-sided platforms. A further differentiating factor relates to the type of vulnerabilities disseminated through bug bounty platforms.

Bug bounties vary in terms of the types of vulnerabilities typically disseminated. The hierarchy is usually clear in terms of both extrinsic and intrinsic rewards: remote code execution vulnerabilities and related memory corruption issues are usually at the top and web vulnerabilities at the bottom. Like with vulnerability markets in general \cite{Algarni14}, this hierarchy also influences the types of hackers who are likely to successfully discover and disclose vulnerabilities. A bug bounty that targets memory corruption vulnerabilities is likely to attract high-skill professionals, whereas web vulnerabilities are easy to discover even with moderate computing knowledge. This skill gap likely contributes to the typical problems affecting vulnerability disclosure, including the frequent delays for vendors to release patches to the security issues disclosed. The absence of vendors' participation and the context of web vulnerabilities are important characteristics of the one-sided OBB platform.

\section{A preliminary analysis of a one-sided platform for web vulnerabilities}\label{section: results}

In what follows, a preliminary empirical analysis is presented about the one-sided OBB platform based on a dataset collected from the platform's online website in October 2017. The dataset analyzed contains $158794$ web vulnerabilities. Before proceeding to tackle this vast amount with descriptive statistics, a brief discussion is necessary about the type of vulnerabilities present in the dataset and the vulnerability disclosure practices on the OBB platform. After this discussion, the empirical analysis proceeds by first considering the evolutionary and productivity aspects in relation to earlier work done by Zhao, Grossklags, and associates (see \cite{ZhaoGrossklags15} in particular). The second part of the empirical analysis focuses on the evaluation of new submissions and the time vendors (or, rather, websites) take to patch the issues disclosed to them.

\subsection{Web Vulnerabilities and the OBB Platform}

The OBB platform only permits the submission of cross-site scripting (XSS) and cross-site request forgery (CSRF) vulnerabilities. Ever since the initial surge in the early 2000s, cross-site scripting bugs have continued to remain the likely most common software vulnerability type~\cite{Kuhn17, Scholte12}. Although there exists a myriad of different XSS vulnerabilities, the essence is that an attacker injects malicious code to a benign website, and this code is executed by the browser of another user visiting the website (see~\cite{GuptaGupta17} for a concise recent review on XSS). Even though automated black box scanners are presumably used also in the bug bounty context~\cite{Algarni14, Bazzoli16, LiDas14}, cross-site scripting vulnerabilities are also easy to discover simply by browsing a website and inserting small blocks of code to parts that require user input.  In contrast to XSS, a CSRF vulnerability exposes a weakness that may trigger a client to make an unintended request (for the background see, e.g.,~\cite{Czeskis13}). For instance: if a client is currently authenticated to a website, a CSRF vulnerability may be exploited by luring the client to click a forged link that makes an unwanted state change, such as changing the client's password. 

In general, cross-site scripting vulnerabilities are much more common than CSRF vulnerabilities. This point can be shown also with the OBB dataset: only as little as $38$ of the vulnerabilities in the dataset dealt with CSRF. Thus, OBB is very much a platform specialized to cross-site scripting vulnerabilities, just like its predecessor, XSSPosed, was explicitly. 

The focus on XSS and CSRF vulnerabilities affects the types of hackers who are likely to participate on the OBB platform, but there are a couple of additional reasons why the web context is important. First, the discovery of these web vulnerabilities requires no intrusive testing techniques \cite{Stock16}. This point is important for a one-sided platform; reporting vulnerabilities found by intrusive techniques may easily expose a platform to liability questions---after all, legal issues have not been unheard-of also in the web vulnerability context \cite{McKinney08, ZhaoLaszka17}. Likewise, the difficult ethical questions that surround some bug bounties~\cite{Egelman13, Wolf16} are a lesser concern in the OBB case. Second, the dynamics of finding vulnerabilities are different. The world wide web is an endless resource. By implication and in contrast to some vendor-specific bug bounties~\cite{Maillart17, ZhaoLaszka17}, it cannot be assumed that finding new vulnerabilities would get more difficult over time in the OBB case. Given that the current size of the indexed world wide web is estimated to be around five billion pages or more~\cite{vandenBosch16}, there will always be vulnerable websites to discover---and rediscover. 

\subsection{Disclosure on the OBB Platform}\label{subsec: disclosure}

\begin{figure}[th!b]
\centering
\includegraphics[width=8cm, height=5cm]{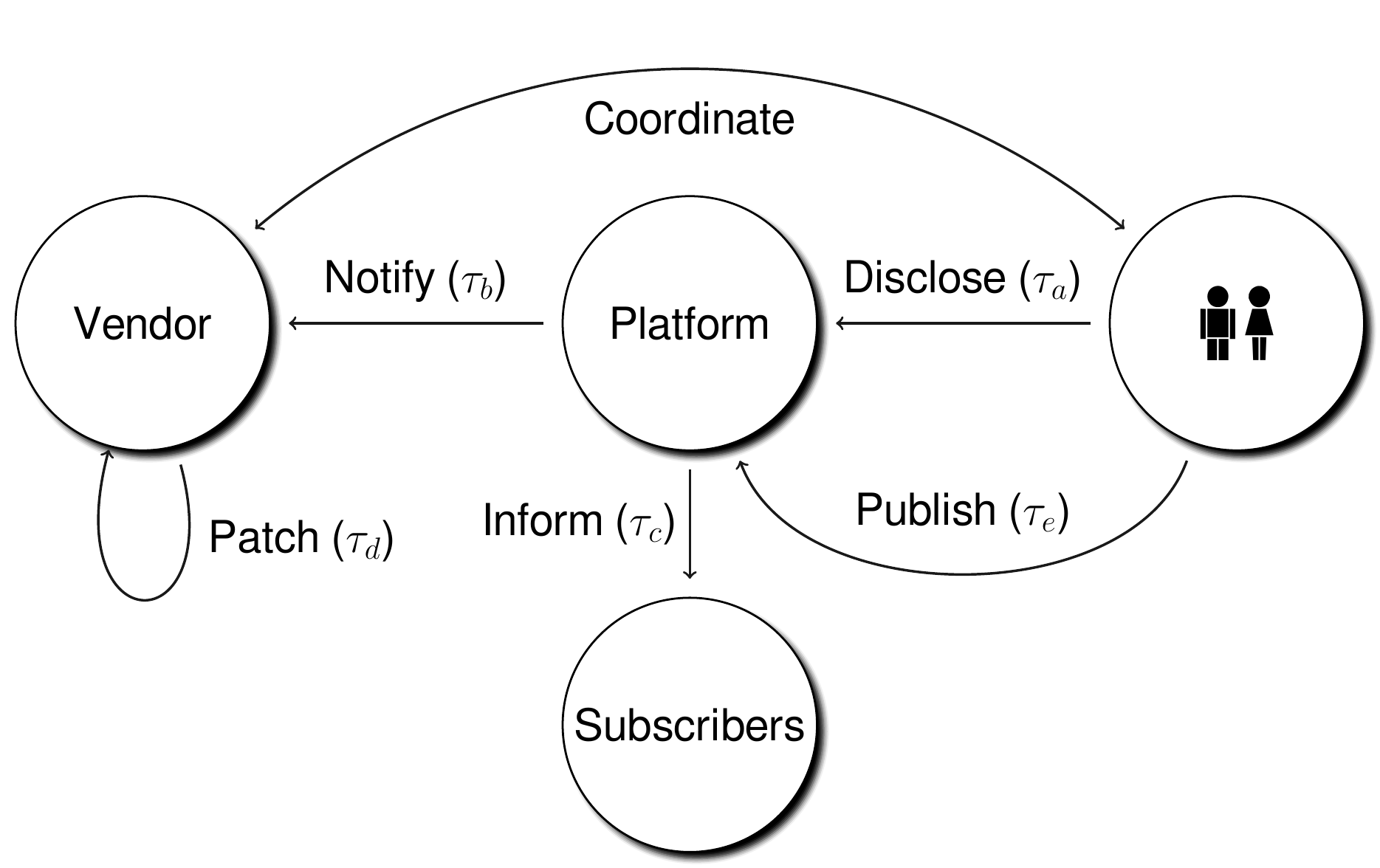}
\caption{Disclosure on the OBB Platform (analytical sketch)}
\label{fig: disclosure}
\end{figure}

The OBB's disclosure model is illustrated in Fig.~\ref{fig: disclosure} with four abstract actor types and six events. Five of the events equate to timestamps marked with a symbol $\tau$. Various axioms could be postulated for the possible ordering of these events, but, in general, it suffices to note that at least $\tau_a \leq \tau_b \leq \tau_d$ always holds, provided that all three events actually occur. While $\tau_a$ and $\tau_b$ are always defined in the analytical model, the set $\lbrace \tau_c, \tau_d, \tau_e \rbrace$  may be undefined. For instance, the timestamp $\tau_d$ remains undefined in case a vendor never patches its website.
 
The process starts when a hacker discloses a vulnerability to the platform ($\tau_a$). As is typical in the vulnerability context~\text{\cite{Ruohonen17TIR, Ruohonen16RCIS}}, the platform orchestrator then carries out an evaluation to rule out fake submissions, bugs that are not actually vulnerabilities, and other false positives that are typical menaces of bug bounties \cite{Laszka16, ZhaoLaszka17}. After the evaluation has been completed, either a customized message or a bulk notification is send to the vendor~($\tau_b$). This notification includes the contact details for reaching the hacker who made the discovery and disclosed the vulnerability on the platform. 

Thus, 
\begin{equation}\label{eq: tte}
\TTE = \tau_b - \tau_a \geq 0 
 \end{equation}
defines a straightforward \textit{time-to-evaluate} (TTE) metric. The actual coordination is left for the two parties. If a vendor acknowledges the notification, either explicitly or implicitly, and the possibly required further coordination is successful, the vendor likely also patches the vulnerability at time $\tau_d$. Provided that $\tau_d$ is defined, such that patching occurred, a conventional metric \cite{Jo17}, say \textit{time-to-patch} (TTP), is given by
\begin{equation}\label{eq: ttp}
\TTP = \tau_d - \tau_b \geq 0 .
\end{equation}

The model also includes a set of subscribers for whom the OBB platform provides security information about the vulnerabilities disclosed through the platform. Moreover, the public disclosure on the platform occurs at $\tau_e$ based on the discretion of a hacker. This assumption is theoretically important. An intermediate actor may hold the decision to publish vulnerability information in strong brokerage models~\cite{KannanTelang05}, but OBB leaves this decision to hackers. If a hacker decides not to publish details, $\tau_e$ remains undefined. Given that many of the intrinsic rewards depend on the availability of public information, these cases are supposedly rare, however. 

Finally, it should be noted that OBB follows the so-called responsible disclosure~\cite{Ruohonen16AICCSA} by providing a grace period before hackers are allowed to disclose public information on the platform. If a vendor patched a vulnerability, a $30$ day grace period is provided; otherwise a $90$ day restraint is applied \cite{OBB17}. These lengths are comparable to current industry practices. 

\subsection{Evolution}

\begin{figure}[th!b]
\centering
\includegraphics[width=\linewidth, height=11cm]{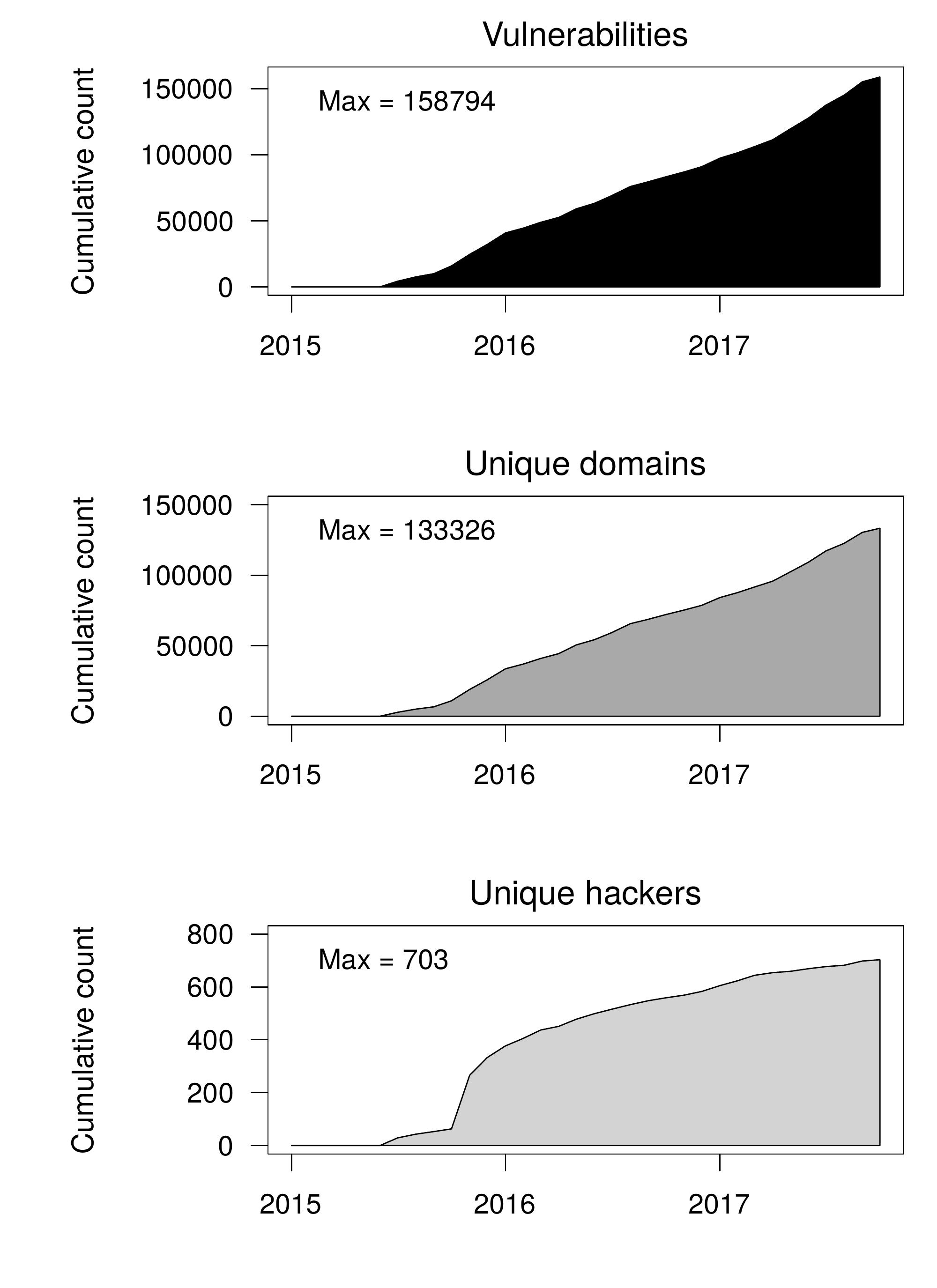}
\caption{Three Key Metrics on the Initial Evolution of the OBB Platform}
\label{fig: evolution}
\end{figure}

The initial evolution of a new platform provides a good bird's-eye view on the prospects for the future success of the platform. The volume of vulnerabilities disseminated through a bug bounty platform provides the most straightforward metric to observe the evolution \cite{ZhaoGrossklags15}. In addition, it is important to consider the network effects; reaching a critical mass is important early on. For a new two-sided bug bounty platform, it is particularly important to promptly attract a group of well-known vendors. As it is a common mistake to overemphasize pricing aspects during the initial evolution of a platform \cite{Casey12}, the names of the vendors likely carry more weight than the compensations paid by the vendors in the early stages. For both two-sided and one-sided bug bounty platforms, a key factor is also the initial amount of active hackers; when a platform is able to attract a group of productive hackers early on, the incentives may intensify for others to join. This direct network effect is fostered by social media. Analogous to other bug bounty platforms~\cite{HuangLiu16}, also OBB relies heavily on social media---in fact, a Twitter account is required for reporting vulnerabilities on the platform. Given these basic premises for network effects, Fig.~\ref{fig: evolution} shows three key metrics on the monthly evolution of the OBB platform during the period observed.

The volume of vulnerabilities disseminated and the amount of unique domains that were affected by the vulnerabilities both follow almost a perfectly linear trend. A peek at the \text{$y$-axes} also reveals that only one XSS vulnerability was reported for most of the domains. In fact, only about 26\% of the unique domains were affected by multiple vulnerabilities. This observation is interesting because it would seem sensible to hypothesize that finding one XSS vulnerability would increase the probability of finding more. Given the social aspects of bug bounties (see Subsection~\ref{subsec: rewards}), it could be also asserted that when one hacker finds web vulnerabilities from a domain, the probability would increase for others to also take a look at the domain. This line of reasoning does not seem plausible according to the results, however. As will be discussed later on, the explanation likely relates to the use of automated tools.

The amount of unique hackers participating on the OBB platform started to rapidly increase in the late 2015. After this initial surge, the growth rate in the amount of new participants has slowly started to decrease. Although the period observed is too short for making definite conclusions, this mild deceleration hints that there may be a saturation point in terms of the global amount of hackers who are engaging with bug bounties. Thus, a~good hypothesis for further work would be the examination of a potential S-shaped growth curve that typically characterizes the diffusion dynamics of many online platforms in general~\text{\cite{Casey12, OndrusLyytinen15}}. Another point is that the total amount of unique participants generally aligns rather well with previous empirical observations about bug bounties. 

Although the total of $703$ unique hackers is much lower than what has been observed for Chinese bug bounties~\text{\cite{HuangLiu16, ZhaoGrossklags14}}, the amount is still comparable in magnitude to platforms such as HackerOne~\cite{ZhaoGrossklags15}. There likely exists also a crossover effect; hackers tend to switch from one bug bounty to another~\cite{Maillart17}. Such crossover effects were typical already in the older vulnerability markets. For instance, a common speculation has been the possibility to sell a vulnerability in one market and an exploit for the vulnerability in another \cite{Ruohonen16RCIS}. Likewise: direct, full, or some other type of vulnerability disclosure may be pursued only after a failed attempt to sell the vulnerability to a broker. An analogous pattern may be present with respect to bug bounties: if a hacker failed to obtain a compensation from a two-sided bug bounty platform, she may decide to publish the discovery on a community-based  platform such as OBB. Duplicate reports between platforms are also a real possibility.

\subsection{Productivity}

The little over seven hundred hackers who participated on the OBB platform between 2015 and late 2017 disclosed nearly $160$ thousand XSS vulnerabilities. This amount is substantial even when keeping in mind the low-profile of cross-site scripting bugs. The volume is also so large that the discoveries cannot have had happened through manual inspection of websites. In other words, human intelligence is generally important for finding security bugs~\cite{Maillart17}, but human intelligence may be even more important in terms of engineering software solutions for automated (web) vulnerability discovery. Manual source code inspection often works well for finding new vulnerabilities~\cite{FangHafiz14, Finifter11}, but automatic scanning is also a necessity insofar as the whole world wide web is the target. Although finding XSS bugs does not require extensive knowledge as such, engineering automated scanners is another thing. The resulting knowledge gap is one factor contributing to the typically uneven distribution of web vulnerability disclosures among hackers participating on a bug bounty platform. 

For instance, the most productive hacker on the OBB platform has disclosed over $23$ thousand XSS vulnerabilities. This amount is comparable in magnitude to what can be reached with large-scale Internet scanning of cross-site scripting vulnerabilities~\cite{Lekies13}, especially since many websites continue to remain vulnerable even after the corresponding web vulnerabilities have been disclosed and reported \cite{GuptaGupta17, Stock16}. However, the overall consequences for a bug bounty platform presumably remain similar irrespective of whether the vulnerability discoveries result from manual labor or automated tools. Analogous to many online platforms in general, a long-tailed probability distribution likely follows in terms of the per-hacker amount of vulnerability disclosures made on a platform.

\begin{figure}[th!b]
\centering
\includegraphics[width=\linewidth, height=5cm]{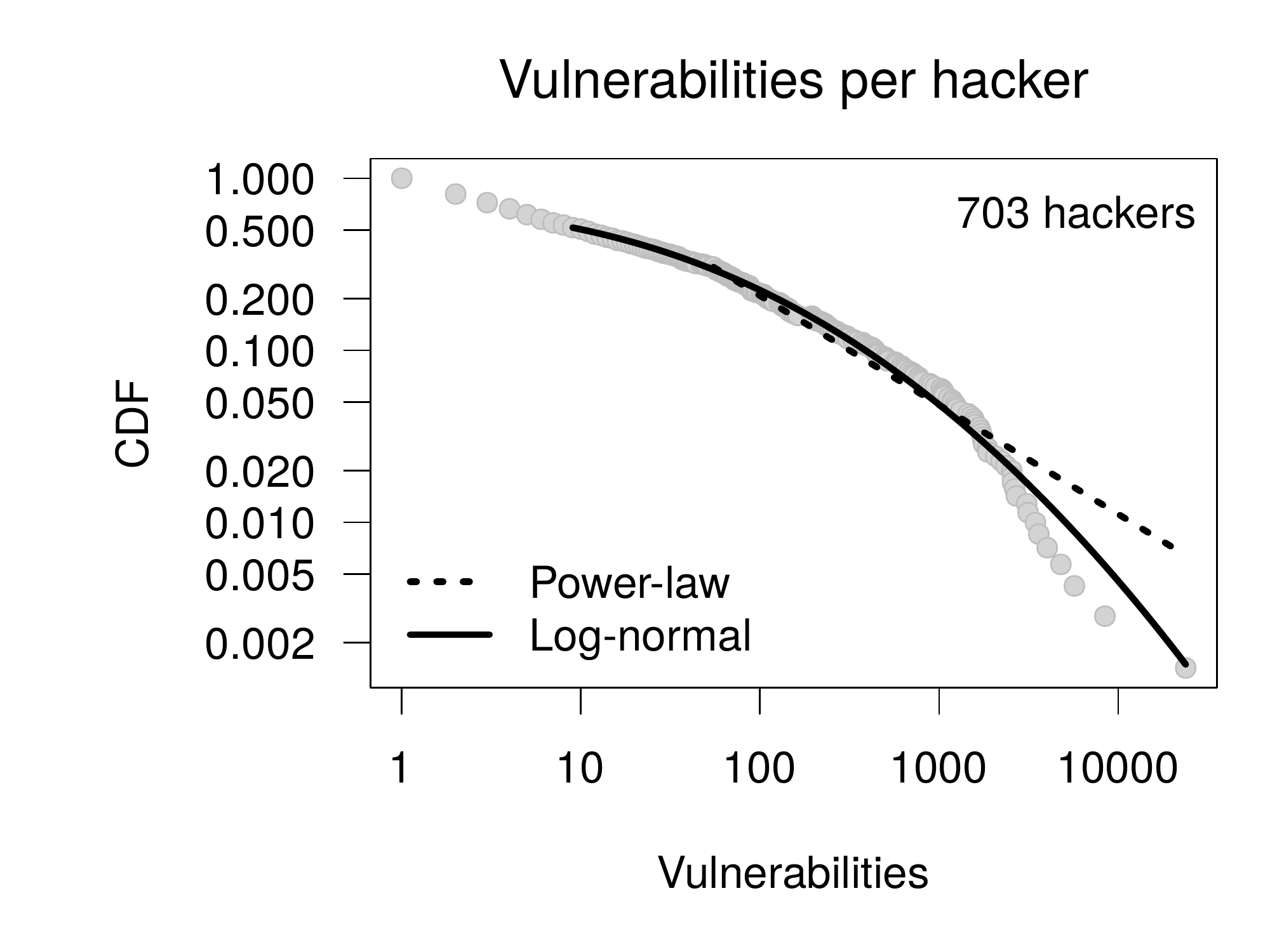}
\caption{Productivity of Hackers}
\label{fig: productivity}
\end{figure}

To examine this typical productivity gap further, Fig.~\ref{fig: productivity} displays the cumulative distribution function (CDF) for the per-hacker amount of vulnerabilities disclosed on the OBB platform. The two distribution approximations visualized are based on the classical estimation setup for examining so-called power-laws \cite{Clauset09, Gillespie15}. In addition to the apparent productivity gap, there are three points worth making from the illustration.

First, the log-normal distribution seems to provide a slightly better fit than the power-law one. From a purely empirical point of view, this stylized fact is noteworthy because it differs from previous observations about bug bounties \cite{Maillart17, ZhaoGrossklags14}. The log-normal distribution is also more challenging to interpret theoretically than a power-law distribution that can be attached to theoretical constructs such as preferential attachment.

Second, the productivity gap causes a vulnerability for the OBB platform. The same applies to most bug bounties \cite{ZhaoGrossklags15}. If the OBB platform would lose some of the most productive hackers, the volume of new vulnerabilities disseminated would presumably decrease substantially. This risk is fostered by the network effects that work also toward the reverse direction. In other words: when a sufficiently large group of participants abandon a platform, a second group of participants may follow. In addition to providing incentives for the most productive hackers to stay on board, it is important to consider means by which the productivity of other hackers could be improved. Given that about 19\% of the unique hackers on the OBB platform have disclosed just one vulnerability, which is fairly typical for bug bounties \cite{Hata17}, a further challenge relates to the common question of how to transform the apparent one-shot visitors into persistent users of the bug bounty platform.

Third, the uneven productivity has also other important theoretical consequences. In particular, the gap implies that merely increasing the volume of hackers is unlikely to substantially increase the volume of vulnerabilities disseminated. This assumption runs in counter to the basic hypotheses often made in the general platform literature~\cite{Voigt16, Finifter13}. 
In the context of web vulnerabilities, an analogous diversity tenet can be also approached in terms of the websites and domains affected. 

\begin{figure}[th!b]
\centering
\includegraphics[width=\linewidth, height=10cm]{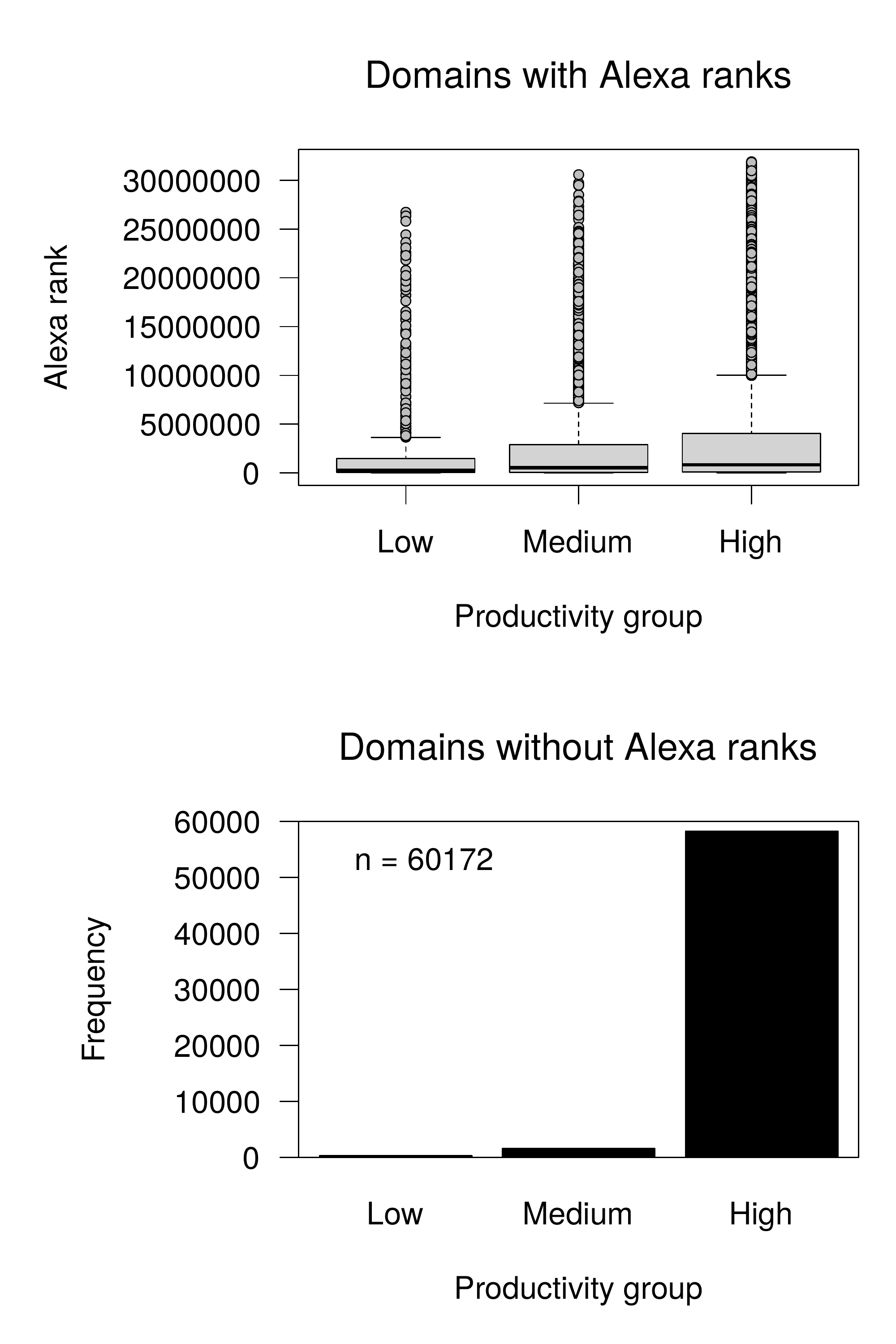}
\caption{Productivity and Popularity of Websites}
\label{fig: prodpop}
\end{figure}

The popularity of the domains offers a simple way to probe this kind of diversity. The basic hypothesis is that hackers who are particularly productive would target more popular domains and websites hosted from these domains, while the rank and file hackers would focus on the less popular domains~\cite{ZhaoGrossklags15}. The basic idea is sound. For instance, there is still some prestige involved in discovering XSS vulnerabilities from websites owned by Google or Netflix---or from online banking sites for which even XSS can pose a real threat. Following existing research \cite{ZhaoGrossklags14, ZhaoGrossklags15}, the hypothesis can be examined by plotting Alexa's popularity ranks against different productivity groups. The three groups used in Fig.~\ref{fig: prodpop} are based on a simple classification: the ``low'' productivity group refers to hackers who have disclosed less than or equal to the median of ten vulnerabilities per-hacker, the ``high'' productivity group contains those who have disclosed more than the 75th percentile of the per-hacker disclosures made, and the ``medium'' group sets in-between these two groups. The popularity ranks refer to those given by the OBB platform.

The results are more or less consistent with previous observations \cite{ZhaoGrossklags15}. Although there is hardly a difference between the median popularity ranks of the three groups, there is a small tendency for the highly productivity hackers to disclose XSS vulnerabilities affecting less popular domains. This observation is reinforced by the lower plot in Fig.~\ref{fig: prodpop}, which shows that almost all of the vulnerabilities affecting unpopular domains without Alexa's ranks were disclosed by the high-productivity group. Thus, all in all, the diversity-based hypotheses do not seem sensible for the OBB platform. The explanation may again relate to automation. In other words, running a large-scale XSS scanner is dependent on the empirical sampling and seeding characteristics. If a scanner uses hyperlinks and web crawling to find new targets, it is unlikely that the cross-site scripting findings would be consistent with website popularity.

\subsection{Evaluation}

The evaluation of new submissions is a generic problem in bug bounties and software bug tracking in general. The problems in triaging of vulnerability reports have also intensified in recent years, partially owing to the popularity of bug bounties and their monetary compensations that tend to incentivize poorly assembled reports and even fake submissions \cite{Laszka16, Ruohonen17TIR}. Given the fundamental nature of the problem, there is also a long history in software engineering for automating at least some aspects of bug triaging~\cite{Uddin17}. The volume of web vulnerabilities disseminated through the OBB platform implies that automation is also a necessity. Fortunately, automatic evaluation of typical cross-site scripting vulnerabilities is easy.

\begin{figure}[th!b]
\centering
\includegraphics[width=\linewidth, height=4.2cm]{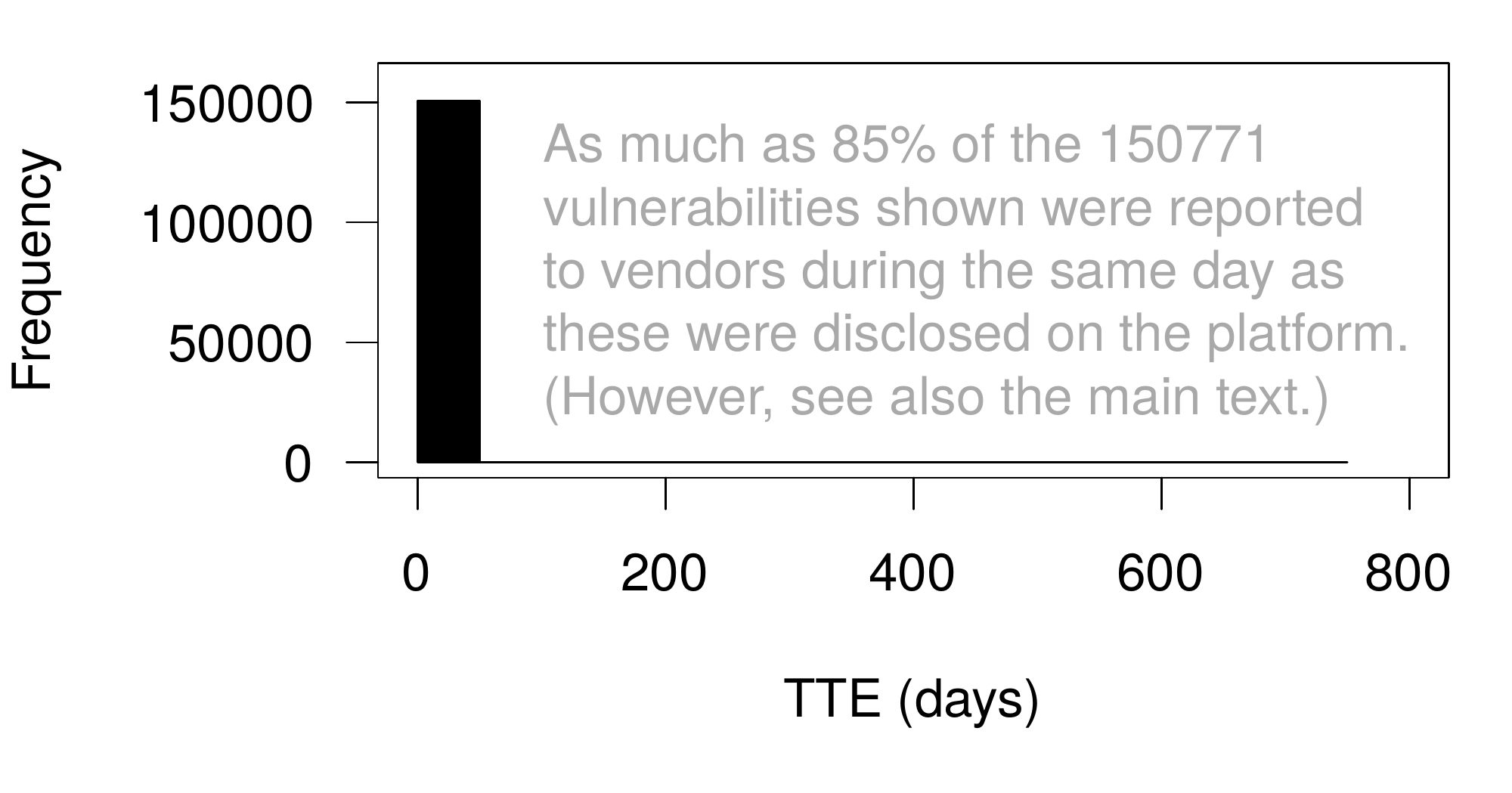}
\caption{Evaluation Times}
\label{fig: tte}
\end{figure}

The OBB platform uses a simple web form for reporting new vulnerabilities. For XSS issues, the form contains the typical \texttt{<script>alert('XSS')</script>} -style payload embedded to a uniform resource locator together with potentially required parameters. Given the information submitted via the form, automatic verification is easy. It should be also possible to use polling for automatically evaluating whether and when the issue is patched by the website affected~\cite{Stock16}. Given this background, the results summarized in Fig.~\ref{fig: tte} are understandable and sensible. Before continuing further, it should be remarked that about eight thousand vulnerabilities had to be removed due to missing $\tau_a$ or $\tau_b$ used to define TTE. Furthermore, the sketch in Fig.~\ref{fig: disclosure} is not entirely accurate because some older vulnerabilities are rather accompanied with explicit timestamps denoting the dates and times on which the vulnerabilities were evaluated. When computing the TTE metric, these explicit timestamps are used when available. Furthermore, newer vulnerabilities are actually accompanied with two notification timestamps that record the dates and times on which custom and generic notifications were sent to the vendors. The smaller of the these is used to define $\tau_b$, provided that an explicit evaluation timestamp is not available.

Thus, the OBB platform is good at automatic evaluation of new submission due to the focus on XSS. However, there is another viewpoint to the time-to-evaluate metric. As was discussed in Subsection~\ref{subsec: disclosure}, the timestamp $\tau_b$ refers to a vulnerability notification sent to a vendor. While the evaluation of a XSS vulnerability may be fast, establishing a contact to the website or domain affected is an entirely different thing. In fact, previous research indicates that many responsible vendor-side parties are not even reachable with notifications about web vulnerabilities~\cite{Stock16, vanEeten17}. Analogous point can be made also with the disclosure help offered in some bug bounties~\cite{HackerOne17a}. Thus, either the OBB's orchestrators are exceptionally good at contacting vendors or the $\tau_b$ event in Fig.~\ref{fig: disclosure} is exposed to some validity concerns. In other words, it may well be that the XSS vulnerabilities disclosed on the platform are verified to be real, but the contacts made to vendors contain shortcomings. This potential deficiency contributes directly to the times vendors take to patch the cross-site scripting issues reported to them.

\subsection{Patching}

The time a vendor takes to patch its products is a classical research topic in the vulnerability disclosure literature. In addition to the noted reporting aspects, there are numerous factors that may influence the typically lengthy time delays. Among these are the type and severity of the vulnerabilities disclosed, the products affected and their age, the quality of a software source code base in general, shared code bases and third-party libraries, company policies and governmental regulations, the potential presence of a third-party coordinator and grace periods, trust, communication skills and the ego of a hacker, and numerous related factors \cite{Finifter13, CERT17, Jo17, KannanTelang05, Ruohonen16AICCSA, Ruohonen17COMSIS}. Like with bug fixing in general~\cite{ElMezouar17}, also social media has recently brought a new element that may affect the delays.

While most of these factors may affect the patching times also in the bug bounty context, it seems reasonable to start from the premise that the type and reputation of a bug bounty platform play decisive roles. If a vendor participates in a two-sided platform as a paying customer, the patching times should be faster compared to one-sided and community-based platforms. Another decisive factor would be the vulnerabilities disseminated through a bug bounty. Accordingly, XSS vulnerabilities should be patched relatively fast already because it is usually trivial to correct these in the software source code. This hypothesis does not hold for the OBB case, however.

\begin{figure}[th!b]
\centering
\includegraphics[width=\linewidth, height=5cm]{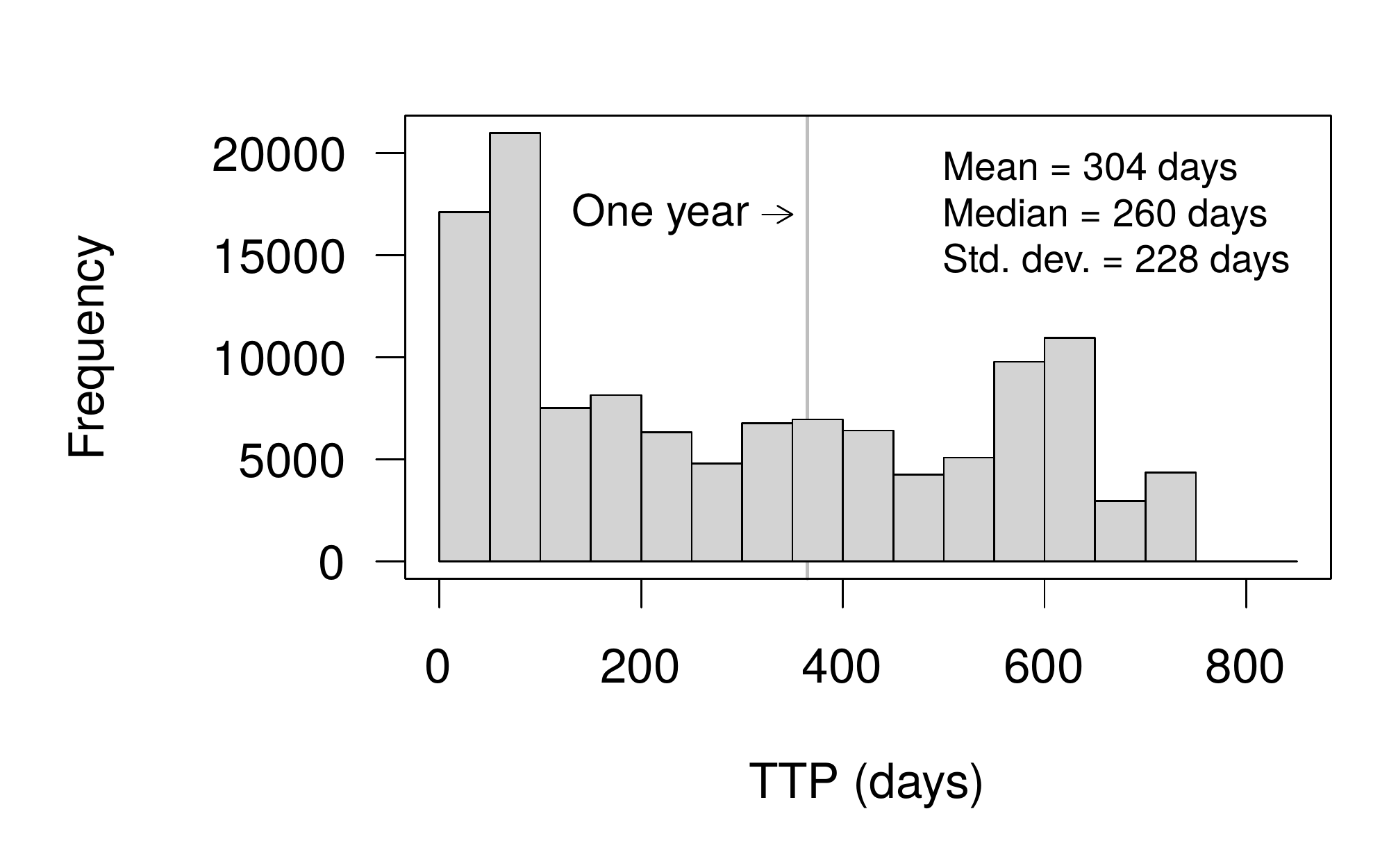}
\caption{Patching Times}
\label{fig: ttp}
\end{figure}

After again removing a little over $32$ thousand vulnerabilities due to missing timestamps, the time-to-patch values can be summarized in the form of a histogram shown in Fig.~\ref{fig: ttp}. As can be seen, the patching times have generally been relatively long. The mean and median of the TTP metric both indicate that vendors have taken over seven months to patch the reported XSS vulnerabilities on average. While the standard deviation is also large, the shape of the histogram does not resemble a long-tailed probability distribution seen for the TTE metric. About 14\% of the vulnerabilities were fixed in less than three months, but rest of the vulnerabilities scatter rather evenly up to the maximum of a little over two years. 

\begin{figure}[th!b]
\centering
\includegraphics[width=\linewidth, height=6cm]{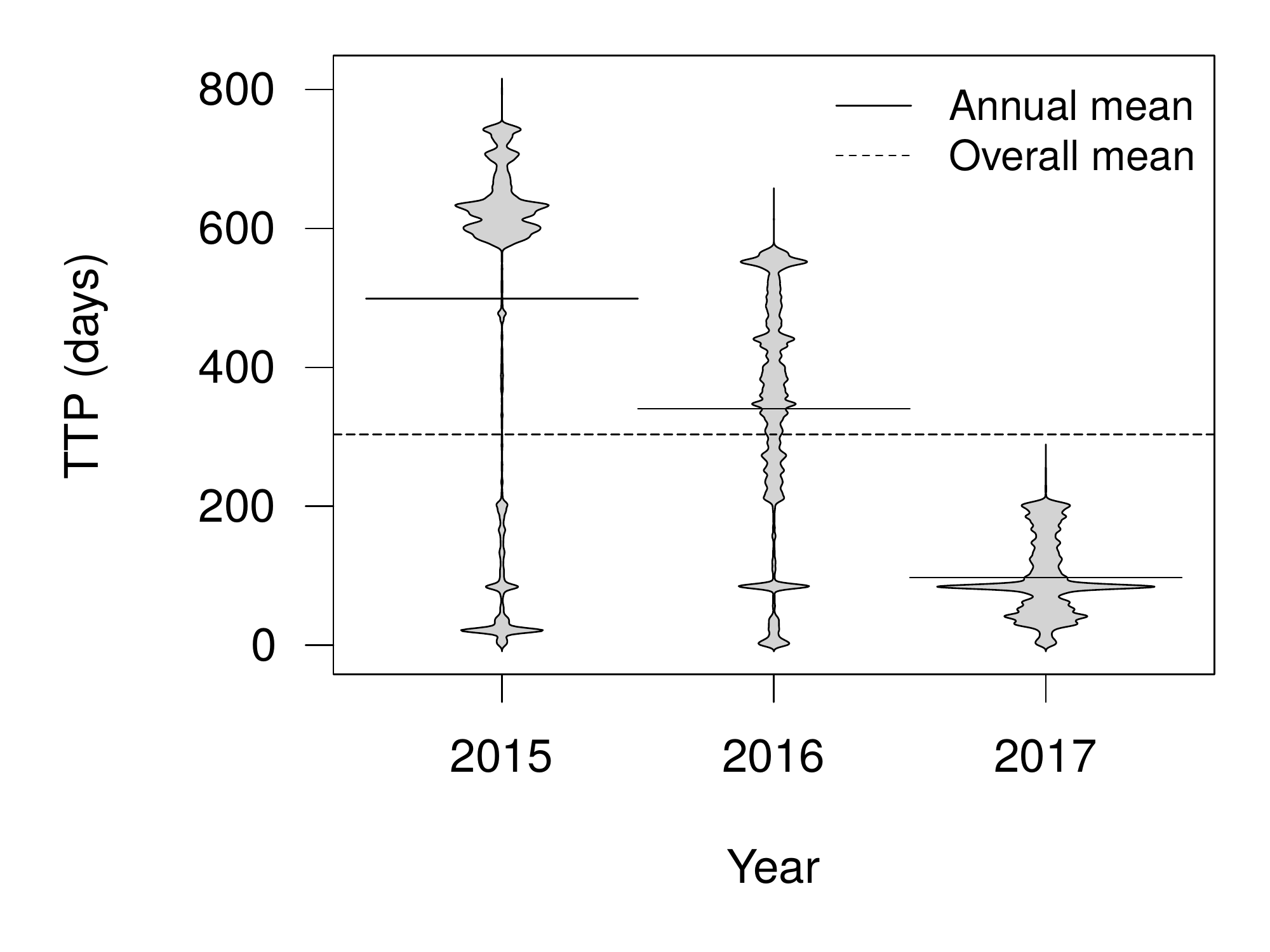}
\caption{Annual Patching Times}
\label{fig: ttp years}
\end{figure}

Despite of the stable linear growth trend in the volume of vulnerabilities disseminated (see Fig.~\ref{fig: evolution}), the patching times have also steadily decreased after the initial launch of the platform (see Fig.~\ref{fig: ttp years}). A possible explanation may be simple: OBB has become more popular and better known already due to the volume of vulnerabilities disseminated. Another point is that the TTP values are not bad as such. In fact, the patching times are even surprisingly similar to those that have been observed for bug bounties targeting conventional software products such as web browsers~\cite{Finifter13}. Of course, this remark should not be used to hide the fact that many of the websites and domains affected remain vulnerable. The missing values for the $\tau_d$ timestamps hint that at least 19\% of the cross-site scripting vulnerabilities are likely still exploitable today.

\begin{table}[th!b]
\begin{center}
\caption{Correlations Between Popularity and Delay Metrics}
\label{tab: cor}
\begin{tabular}{lccccr}
\toprule
&\qquad& Time-to-evaluate &\qquad& Time-to-patch & \qquad\\
\cmidrule{3-3}\cmidrule{5-5}
Subset $n$ && $94145$ && ~~$75834$ \\
Pearson $r$ && $0.002$ && $-0.003$ \\
\bottomrule
\end{tabular}
\end{center}
\end{table}

\begin{figure}[th!b]
\centering
\includegraphics[width=\linewidth, height=8cm]{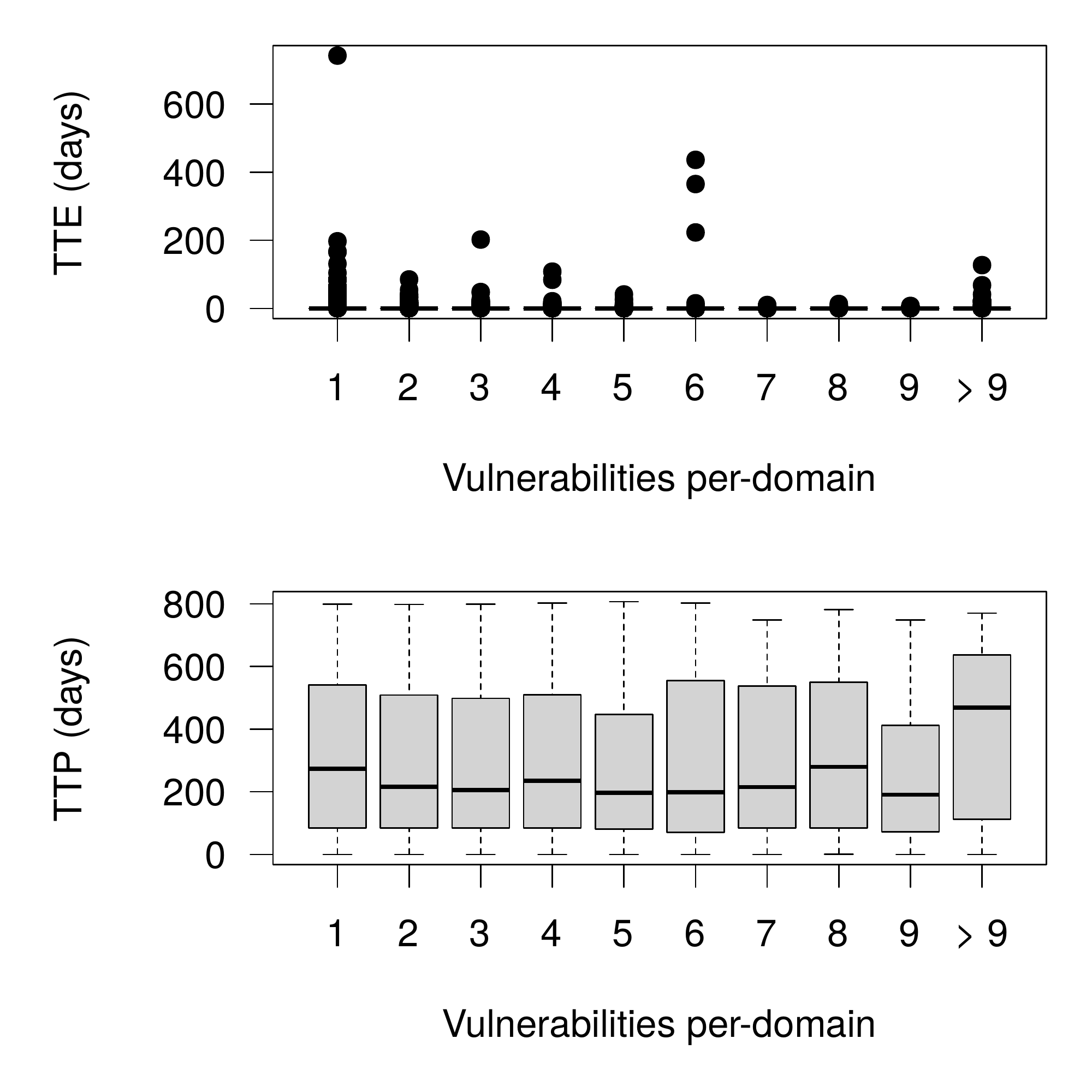}
\caption{TTE and TTP Across Domains Affected by Multiple Vulnerabilities}
\label{fig: tte ttp domains}
\end{figure}

\begin{figure}[th!b]
\centering
\includegraphics[width=\linewidth, height=4.5cm]{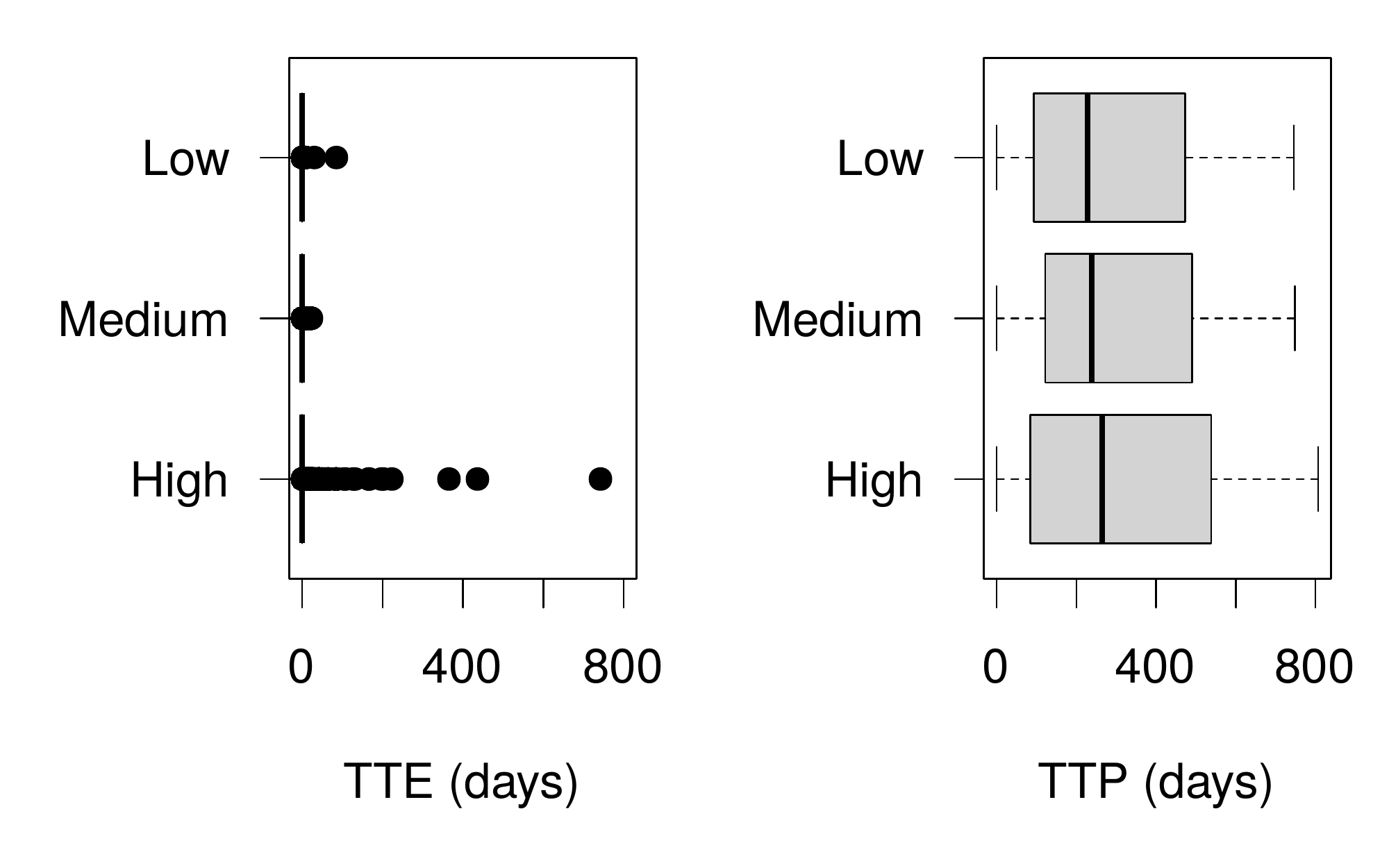}
\caption{TTE and TTP Across Hackers' Productivity Groups}
\label{fig: tte ttp prod}
\end{figure}

There are a couple of additional hypotheses worth briefly visiting. The first is that the popularity of a domain would proxy the maintenance effort for the domain, and this effort would correlate with the patching times. As there are multiple dimensions to the concept of effort in the web context~\cite{Tajalizadehkhoob17}, an alternative speculation could be that the evaluation and patching times would be shorter for the popular domains due to the availability of unambiguous contact details. Regardless of the interpretation, the numbers shown in Table~\ref{tab: cor} indicate no relation between the popularity ranks and the delay metrics.

The other hypothesis relates to learning effects. The basic theoretical rationale for such effects is simple: a hacker who discloses thousands of vulnerabilities may become better at communicating with vendors, while a domain affected by multiple vulnerabilities may patch faster due to a similar learning effect. According to Fig.~\ref{fig: tte ttp domains}, the medians for the TTE and TTP do not notably differ between domains affected by a single XSS and domains exposed to multiple cross-site scripting bugs. Nor is there a visible learning effect in terms of the three productivity groups for the hackers. In general, learning from the past seems limited on the OBB platform.

\section{Discussion}\label{section: discussion}

The results presented allow to start with an answer to the always good question about whether things might have changed regarding basic programming mistakes such as input validation~\cite{Scholte12}. Alas, the answer is clearly negative. Despite of over a decade worth of education and security awareness campaigns, basic issues such as cross-site scripting are still extremely common. In fact, the nearly $160$ thousand web vulnerabilities disclosed through the OBB platform between 2015 and late 2017 is about twice the whole amount currently archived to the National Vulnerability Database, for instance.

The initial evolution of the OBB platform indicates that there is at least a good promise for bug bounty platforms that do not provide explicit monetary compensations. The amount of XSS vulnerabilities disclosed through the OBB platform is a laudable achievement from the idealistic viewpoint that the fundamental goal would be to improve security in the world wide web. The platform has also made its own contribution to security awareness through the communication with website maintainers and web developers, as well as through the presence in Twitter. There are also many challenges for both research and practice. To outline some of these challenges, the remainder of the paper enumerates a few directions for further research and catalogs some points that may be worthwhile to consider by the orchestrators of current bug bounty platforms.

The presented distinction between one-sided and two-sided bug bounty platforms offers a good way to proceed into more theoretically motivated comparisons. Given the continuing lack of business model research in the contexts of vulnerability markets and security industry in general \cite{Ruohonen16RCIS}, the platform (or ecosystem) literature provides also an extensive base for reference theories, hypotheses, and practical insights. The basic premise here is that there should be a difference between one-sided and two-sided platforms, money and non-money platforms, vendor-specific and community-based platforms, or targeted and untargeted bug bounties. While the contemporary platform economy is very much also a copycat economy, it is still surprising that the basic premise does not clearly manifest itself in the current bug bounties. When comparing the results presented to existing research---and when looking at the explicit comparisons already done in existing empirical research~\cite{ZhaoGrossklags14, ZhaoGrossklags15, Finifter13}, the current bug bounty platforms appear extremely similar to each other in many respects. All typical traits of online platforms and their communities are present. 

Among these traits is the productivity gap between hackers. A small group of people usually do most of the work on online platforms, and a small group of hackers disclose most of the vulnerabilities on bug bounty platforms. The commonplace network effects pose a double challenge for one-sided platforms: a platform should at the same time lure new productive participants and ensure that the existing productive participants continue to use the platform. While both challenges have been implicitly addressed in the existing bug bounty research, there is still room for further empirical inquiries also in this regard. For instance, (1) the so-called \textit{diffusion} tradition \cite{Casey12, OndrusLyytinen15, Ruohonen17ELMA} could be followed for examining not the volumes as such but the acceleration, deceleration, and potential saturation aspects related to the volumes. For both Chinese and Western bug bounties, there may be an upper limit in the potential amount of hackers who are able or willing to engage with bug bounties.

The productivity gap correlates with a knowledge gap. Given that some outlying participants on the OBB platform have disclosed tens of thousands of web vulnerabilities, it seems that the possession and use of automated tools contributes to the productivity gap significantly. By implication, there presumably exists also a knowledge gap in terms of the ability to engineer automated XSS scanners. Cross-site scripting vulnerabilities remain at the bottom of the pile in terms of prestige, but it seems that XSS alone can differentiate a hacker demography. Narrowing the productivity gap by narrowing the knowledge gap is an important point for bug bounty orchestrators to consider. Knowledge sharing is at the center of this question. But while learning from others is often noted as important for bug bounties \cite{Ring14, ZhaoGrossklags15}, following someone in social media hardly equates to actual learning. 

Thus, (2) it may be important to encourage the \textit{sharing} of technical details and software source code for finding new web vulnerabilities. Given the competitive aspects intentionally promoted in current bug bounties, voluntary sharing of technical details may be difficult to achieve, however. For balancing the competition, it might be reasonable to recommend that the bug bounty orchestrators should themselves provide blueprints and drafts for web vulnerability discovery \`a la Stack Overflow. After all, ``smashing the stack for fun and profit'' was a decisive historical learning moment for the older generation of hackers.

Now that the OBB platform has geared itself toward subscriptions and multiple types of web vulnerabilities, it becomes important for the orchestrators to consider the means by which quality could be incentivized over quantity. Given the lack of monetary compensations, (3) improving the \textit{gamification} techniques may provide a lever, although it remains unclear how these techniques actually influence the intrinsic rewards. Even when monetary compensations cannot be relied upon, (4) the common \textit{subsidization} techniques \cite{Casey12} may provide a further option. For instance, it might be possible to consider (per hacker or per domain) varying grace periods, or to reward those hackers who are able to successfully coordinate with vendors. From a more contentious viewpoint, (5) even the platform's current \textit{strategic goals} might be challenged. In other words, perhaps it is precisely XSS and the associated large-scale Internet scanning that sets OBB apart from the competitors. After all, platforms such as Shodan have shown that there is a general demand for Internet scanning services.

In terms of Internet scanning, the (open) data provided by the OBB platform leaves also some question marks. In particular, something does not match in terms of the evaluation times presented. While it is possible to automate the verification of XSS bugs, it is a different thing to automate successful notifications sent to the websites and domains affected. The platform's orchestrators have promoted new initiatives such as the so-called~\texttt{security.txt}~ proposal~\cite{Foudil18}, but the problem is that even the older reporting standards are frequently ignored by vendors and websites' owners~\cite{vanEeten17}. In terms of further research, (6)~a good question would be to evaluate how well the \textit{reporting practices} perform in practice. The idea about submitting vulnerability reports based on research results has also been toyed in the past~\cite{Stock16}. The XSS context would make it relatively easy to assemble a sufficient empirical sample that could be submitted to OBB or (and) some related bug bounty.

The empirical aspects can be approached also from a different angle; open data can be a success factor for bug bounty platforms \cite{ZhaoGrossklags15}. More generally, (7) there would be an excellent opportunity particularly for one-sided community platforms to ride on the current artificial intelligence and \textit{data mining} hype. For instance, the results presented indicate that website popularity, the productivity gap, and the learning effects do not affect the patching times. However, (8) the \textit{maintenance effort} of websites would be a good hypothesis to examine in further research \cite{Tajalizadehkhoob17}. To avoid so-called post-hoc analysis, it would be preferable for data mining if a platform would record basic information about websites before and after a vulnerability is reported or patched. In terms of the potential data to record, a good start would be the hypertext transfer protocol headers and the hypertext markup language of the web pages affected. 

In terms of empirical academic research, it can be also noted that the existing bug bounty research (including this paper) has focused on different aggregated metrics such as the volumes of vulnerabilities and hackers. However, (9) it may well be that contextual \textit{outliers} are more interesting and relevant to examine. For instance, the character string \texttt{bank} appears in $464$ domain names present in the sample used in this paper. Also \texttt{equifax.com} and \texttt{equifax.co.uk} are represented.

The points mentioned can be combined with other business considerations. The business models underneath the current bug bounty platforms have followed closely the historical heritage from the older vulnerability markets. However, it seems that some of the historical ideas and options have not yet been tried in the bug bounty context. For instance, (10)~\textit{auctions}, product \textit{bundles}, and \textit{insurance} contracts have been frequently discussed in the context of vulnerability markets~\cite{Bohme06, Miller07, Ruohonen16RCIS}, but concrete experimentation with such options seems limited in the bug bounty context. While such options are more relevant for the two-sided bug bounty platforms, new openings are available also for community-based one-sided platforms. 

Given the popularity of different hacking contests~\cite{Nakaya17} and security exercises \cite{Mullins07}, it might be possible to move away from the prevalent vendor-specific paradigm by considering the subscription concept from a different viewpoint. In terms of the OBB platform and cross-site scripting, (11) one option might be to offer a service for registering \textit{domain name lists} to be scanned and evaluated by the hackers participating on the platform. Such a service would allow to also evaluate whether a platform can actually make a difference in terms of security.

Finally, the current bug bounty ecosystem as a whole can be viewed also as a rivalry between different platforms. The competitive aspects offer also new business opportunities. For instance, to maximize revenues, publicity, and user participation, (12)~\textit{alliances} between platforms are commonly used in some industry sectors~\cite{Casey12, OndrusLyytinen15}. In addition to the maximization aspects, alliances may reduce the frequent hopping from one bounty to another \cite{Maillart17}, which likely contributes to the prevalence of invalid submissions, false positives, duplication of work, and related practical problems. Now that also governmental agencies and public sector organizations have started to participate in bug bounty platforms and implement bug bounties of their own, it becomes important to consider whether CERTs or related institutions should participate in the bug bounty ecosystem. New innovations are required also for improving the vulnerability disclosure practices and processes, which continue to be a bottleneck also in many bug bounties. 


\balance
\bibliographystyle{abbrv}

\end{document}